\documentclass[a4paper]{article} 

\usepackage{graphicx,enumerate,amsmath}

\usepackage{amsmath,amsfonts,amssymb,amsthm}

\newtheorem{teo}{Theorem}
\newtheorem{prop}[teo]{Proposition}

\newtheorem{rmk}{Remark}

\usepackage{color,hyperref}
\definecolor{dark-blue}{rgb}{0.15,0.15,0.4}
\definecolor{dark-red}{rgb}{0.4,0.15,0.15}
\definecolor{medium-red}{rgb}{0.6,0,0}
\definecolor{medium-blue}{rgb}{0,0,0.6}
\hypersetup{
    colorlinks=true,
    linktoc=all,
    citecolor=medium-red,
    filecolor=medium-blue,
    linkcolor=medium-blue,
    urlcolor=medium-blue,
}
        \theoremstyle{plain} 

        \theoremstyle{definition}

        \theoremstyle{remark}
        
\usepackage{color,hyperref}
\definecolor{dark-blue}{rgb}{0.15,0.15,0.4}
\definecolor{dark-red}{rgb}{0.4,0.15,0.15}
\definecolor{medium-red}{rgb}{0.6,0,0}
\definecolor{medium-blue}{rgb}{0,0,0.6}
\hypersetup{
    colorlinks=true,
    linktoc=all,
    citecolor=medium-red,
    filecolor=medium-blue,
    linkcolor=medium-blue,
    urlcolor=medium-blue,
}

                \numberwithin{equation}{section}
\makeatletter
\renewcommand*\env@matrix[1][\arraystretch]{%
  \edef\arraystretch{#1}%
  \hskip -\arraycolsep
  \let\@ifnextchar\new@ifnextchar
  \array{*\c@MaxMatrixCols c}}
\makeatother

\begin{document}
        	
\title{Extensions of non-natural Hamiltonians}

\author{Claudia Maria Chanu, Giovanni Rastelli\\ \\
	Dipartimento di Matematica, Universit\`{a} di Torino,\\ via Carlo Alberto 10, 10123, Torino, Italia\\
	email: claudiamaria.chanu@unito.it; giovanni.rastelli@unito.it}

\date{}

\maketitle

\begin{abstract}
	The concept  of extended Hamiltonian systems allows the geometrical interpretation of several integrable and  superintegrable systems with polynomial first integrals of degree depending on a rational parameter. Until now, the procedure of extension has been applied only in the case of natural Hamiltonians. In this article, we give several examples of application to non-natural Hamiltonians, such as the two point-vortices, the Lotka-Volterra and some quartic in the momenta Hamiltonians, obtaining effectively extended Hamiltonians in some cases and failing in others. We briefly discuss the reasons of these results. 
	\end{abstract}

\section{Introduction} Given a Hamiltonian $L$ with $N$-degrees of freedom, the procedure of extension allows the construction of Hamiltonians $H$ with $(N+1)$ degrees of freedom admitting as first integrals $L$ itself, with all its possible constants of motion, and a characteristic first integral dependent on a rational parameter $k$. We gave several examples of extension of natural Hamiltonians on Riemannian and pseudo-Riemannian manifolds in \cite{CDRPol,CDRfi,CDRgen,CDRsuext,CDRraz,TTWcdr,CDRpw}, including anisotropic, Harmonic oscillators, three-body Calogero and Tremblay-Turbiner-Winternitz systems. In all these examples, $L$ and $H$ are natural Hamiltonians and the characteristic first integral  is polynomial in the momenta of some degree depending on $k$.

However, the procedure of extension do not make any assumption on $L$ other than it is a regular function on some cotangent bundle $T^*M$. Until now, we always considered $L$ as a natural Hamiltonian, in such a way that the extended Hamiltonian is itself natural. In this work, we apply the Extension Procedure on functions $L$ which are no longer quadratic in the momenta and, consequently, the extended Hamiltonian is not a natural one. The construction of an extended Hamiltonian requires the determination of a certain function $G$ well defined  on all $T^*M$, up to some lower-dimensional subset of singular points.  The extended Hamiltonian  is a polynomial in $p_u$, $L$, while its characteristic  first integral is a polynomial in $p_u$, $L$, $G$ and $X_LG$, the  derivative of $G$ with respect to the Hamiltonian vector field of $L$, so that its  global definition depends ultimately on $G$ and $X_LG$. Therefore, our analysis is focused on the determination of the function $G$ in the different cases, and on the study of its global behaviour on $T^*M$. 

Since this work is intended as a preliminary study of the possible applications of the extension procedure to non-natural Hamiltonians, we do not pretend here to obtain complete and general results, but we focus on some meaningful examples only.

In Sec. 2 we recall the fundamentals of the theory of extended Hamiltonians. In Sec. 3 we consider extensions of functions quartic in the momenta and we find examples of extended Hamiltonians in analogy with the quadratic  Hamiltonian case. The analysis becomes more subtle in Sec. 4, when we try to extend functions which are not polynomial in the momenta, as the case of the two point-vortices Hamiltonian. Here the global definition of the extended Hamiltonian and its characteristic first integral becomes an issue in some cases, so that the extension is possible only for some values of the constant of motion $L$, while in other cases the extension is always possible.

We conclude in Sec. 5 with some examples where we are unable to find any properly globally defined extended Hamiltonian.

\section{Extensions of Hamiltonian systems}\label{ex}
Let $L(q^i,p_i)$ be a Hamiltonian with $N$ degrees of freedom, that is defined on the cotangent bundle $T^*M$ of an $N$-dimensional manifold $M$.

We say that $L$ {\em admits extensions}, if there exists $(c,c_0)\in \mathbb R^2
- \{(0,0)\}$ such that there exists a non null solution $G(q^i,p_i)$ of
\begin{equation}\label{e1}
X_L^2(G)=-2(cL+c_0)G,
\end{equation}
where $X_L$ is the Hamiltonian vector field of $L$.

If $L$ admits extensions, then, for any $\gamma(u)$ solution of the ODE
\begin{equation}\label{eqgam}
\gamma'+c\gamma^2+C=0,
\end{equation}
depending on the arbitrary constant parameter 
$C$,	
we say that any Hamiltonian 
$H(u,q^i,p_u,p_i)$ with $N+1$ degrees of
freedom of the form
\begin{equation}\label{Hest}
H=\frac{1}{2} p_u^2-k^2\gamma'L+ k^2c_0\gamma^2
+\frac{\Omega}{\gamma^2},
\qquad k=\frac mn,\, m,n\in \mathbb{N}-\{0\}, \   \Omega\in\mathbb{R}
\end{equation}
is an {\em  extension of $L$}.

Extensions of Hamiltonians where introduced
in \cite{CDRfi} and studied because they admit polynomial in the momenta first integrals generated via a recursive algorithm. 
Moreover, the degree of the first integrals is related with the choice of   $m,n$.
Indeed, for any $m,n\in \mathbb N-\{0\}$, 
let us consider the operator
\begin{equation}\label{Umn}
U_{m,n}=p_u+\frac m{n^2}\gamma X_L.
\end{equation}

\begin{prop}\cite{CDRraz}
	For $\Omega=0$,
	the Hamiltonian (\ref{Hest}) 
	is in involution with the function
	\begin{equation}\label{mn_int}
	K_{m,n}=U_{m,n}^m(G_n)=\left(p_u+\frac{m}{n^2} \gamma(u)   X_L\right)^m(G_n),
	\end{equation}
	where  $G_n$ is the $n$-th term of the recursion
	\begin{equation}\label{rec}
	G_1=G, \qquad G_{n+1}=X_L(G)\,G_n+\frac{1}{n}G\,X_L(G_n), 
	\end{equation}
	starting from any solution $G$ of (\ref{e1}).
\end{prop}


For $\Omega\neq 0$, the recursive construction of a first integral is more complicated:  we construct the following function, depending on two strictly positive integers
$s,r$ 
\begin{equation} \label{ee2}
\bar K_{2s,r}=\left(U_{2s,r}^2+2\Omega \gamma^{-2}\right)^s(G_r),
\end{equation}
where 
the operator $U^2_{2s,r}$ is defined {according to
	(\ref{Umn})} as
$$
U^2_{2s,r}=\left( p_u+\frac{2s}{r^2} \gamma(u)   X_L
\right)^2,
$$
and 
$G_r$ is, as in (\ref{mn_int}),  the $r$-th term of the recursion (\ref{rec}),  
with $G_1=G$ solution of (\ref{e1}).
For $\Omega=0$ the functions (\ref{ee2})
reduce to (\ref{mn_int}) and thus can be computed 
also when the first of the indices is odd.

\begin{teo}\cite{TTWcdr} \label{t2}
	For any $\Omega\in \mathbb{R}$,
	the Hamiltonian (\ref{Hest}) 
	satisfies,
	for  $m=2s$, 
	\begin{equation}\label{cp}
	\{H,\bar K_{m,n}\}=0,
	\end{equation}
	for $m=2s+1$,
	\begin{equation}
	\label{cd}
	\{H ,\bar K_{2m,2n}\}=0.
	\end{equation}
\end{teo}

We call $K$ and $\bar{K}$, of the form (\ref{mn_int}) and
(\ref{ee2}) respectively, \emph{ characteristic first integrals}  of the corresponding extensions.
It is proved in \cite{CDRfi,TTWcdr} that the  characteristic first integrals $K$ or $\bar K$  are functionally independent from $H$, $L$, and from any 
first integral $I(p_i,q^i)$ of $L$.
This means that the extensions of (maximally) superintegrable Hamiltonians	are 
(maximally) superintegrable Hamiltonians with one additional degree of freedom (see also \cite{CDRsuext}).
In particular, any extension of a one-dimensional Hamiltonian is maximally superintegrable.

The explicit expression of the characteristic first integrals is given as follows \cite{CDRraz,TTWcdr}.
For $r\leq m$, 
we have
\begin{equation}\label{EEcal}
U_{m,n}^r(G_n)= P_{m,n,r}G_n+D_{m,n,r}X_{L}(G_n),
\end{equation}
with
$$
P_{m,n,r}=\sum_{ j=0}^{[r/2]}\binom{r}{2 j}\, \left(\frac mn  \gamma \right)^{2 j}p_u^{r-2 j}(-2)^j(cL+c_0)^ j,
$$
$$
D_{m,n,r}=\frac 1{n}\sum_{ j=0}^{[(r-1)/2]}\binom{r}{2 j+1}\, \left(\frac mn  \gamma \right)^{2 j+1}p_u^{r-2 j-1}(-2)^j(cL+c_0)^ j, \quad m>1,
$$
where $[\cdot]$ denotes the integer part and $D_{1,n,1}=\frac 1{n^2} \gamma$.

The expansion of the first integral (\ref{ee2}) is
\begin{equation*}
\bar K_{2m,n}=\sum_{j=0} ^{m}\binom{m}{j}\left(\frac {2\Omega}{\gamma^2}\right)^jU_{2m,n}^{2(m-j)}(G_n),
\end{equation*}
with $U^0_{2m,n}(G_n)=G_n$, and

\begin{equation}\label{Gn_alt}
G_n=\sum_{k=0}^{\left[\frac{n-1}{2}\right]}
\binom{n}{2k+1}(-2(cL+c_0))^k G^{2k+1}(X_L G)^{n-2k-1}.
\end{equation}


\begin{rmk}
	\rm
	In \cite{CDRgen} it is proven that the ODE (\ref{eqgam}) defining $\gamma$ is a necessary condition in order to get a characteristic first integral of the form (\ref{mn_int}) or
	(\ref{ee2}). 
	According to the value of $c$ and $C$, the explicit form of $\gamma(u)$ is given (up to constant translations of $u$) by
	\begin{equation}\label{fgam}
	\gamma= \left\{ 
	\begin{array}{lc}
	-C u & c=0 \\
	\frac{1}{T_\kappa (cu)}=\frac{C_\kappa (cu) }{S_\kappa (cu)} 
	& c\neq 0 			
	\end{array}
	\right.
	\end{equation}
	where $\kappa=C/c$ is the ratio of the constant parameters appearing in (\ref{eqgam}) and $T_\kappa$, $S_\kappa$ and $C_\kappa$  are the trigonometric tagged functions
	(see also \cite{CDRraz} for a summary of their properties)
	$$
	S_\kappa(x)=\left\{\begin{array}{ll}
	\frac{\sin\sqrt{\kappa}x}{\sqrt{\kappa}} & \kappa>0 \\
	x & \kappa=0 \\
	\frac{\sinh\sqrt{|\kappa|}x}{\sqrt{|\kappa|}} & \kappa<0
	\end{array}\right.
	\qquad
	C_\kappa(x)=\left\{\begin{array}{ll}
	\cos\sqrt{\kappa}x & \kappa>0 \\
	1 & \kappa=0 \\
	\cosh\sqrt{|\kappa|}x & \kappa<0
	\end{array}\right.
	$$
	$$
	T_\kappa(x)=\frac {S_\kappa(x)}{C_\kappa(x)}.
	$$
	Therefore, we have 
	\begin{equation}\label{fgamp}
	\gamma'= \left\{ 
	\begin{array}{lc}
	-C  & c=0 \\
	\frac{-c}{S_\kappa ^2 (cu)} 
	& c\neq 0 .
	\end{array}
	\right.
	\end{equation}
\end{rmk}

\begin{rmk} \rm The global definition of the characteristic first integral of the extended Hamiltonian is ultimately determined by the definition of $G$ and its derivative $X_L G$. When these objects are globally defined, then also the characteristic first integral is. 
\end{rmk}

From the brief exposition given above it is clear that the  extensions of a function $L$ on $T^*M$ are completely determined once a solution $G$ of (\ref{e1}) is known, provided it is regular and well defined on $T^*M$. The fundamental step for the application of the extension procedure is therefore the determination of $G$. In all existing examples of extended Hamiltonians the function $L$ is always a quadratic polynomial in the momenta. The examples include the anisotropic harmonic oscillators, the Tremblay-Turbiner-Winternitz and the Post-Winternitz systems. For several of these systems there exist a quantization theory, based on the Kuru-Negro factorization in shift and ladder operators, adapted to Hamiltonians which are extended Hamiltonians \cite{CRsl}.
In order to generalise the extension procedure to non-natural Hamiltonians, we focus our research here on the determination of functions $G$ solution of (\ref{e1}), leaving for other works a deeper analysis of the resulting extended systems. We consider below some examples of non-natural Hamiltonians, or natural Hamiltonians in a non-canonical symplectic  or Poisson structure.
 Since the forms of the extended Hamiltonian and of the characteristic first integral are completely determined once known $L$ and $G$, we are solely concerned with the determination and analysis of $G$ and $X_LG$.

\section{Extensions of Quartic Hamiltonians}

Hamiltonians of degree four in the momenta are considered in \cite{FC}. These Hamiltonians are written in Andoyer projective variables and allow a unified representation of several mechanical systems, such as  harmonic oscillator, Kepler system and rigid body dynamics, corresponding to different choices of parameters. We consider here  some   toy model of Hamiltonians of degree up to four in the momenta.
\begin{enumerate}
	\item Let us assume	
	\begin{equation}
	L=p^4+f_1(q)p^3+f_2(q)p^2+f_3(q)p+V(q).
	\end{equation}
	The exension of $L$ is possible if global solutions $G$ of 
	\begin{equation}\label{LE1}
	X_L^2 G=-2(cL+c_0)G,
	\end{equation}
	are known. If we assume $G(q)$, then the function 
	$$
	G=C_1q+C_2,
	$$
	is a solution of (\ref{LE1})  if $L$ is in the form 
	\begin{eqnarray}\label{sq}
	L=\frac {\left(16C_1 p^2+8C_1fp+2cC_1q^2+4cC_2q+C_1f^2+8C_1C_3\right)^2}{256\,C_1^2}-\frac {c_0}c,
	\end{eqnarray}
	where $C_i$ are real constants and $f(q)$ is an arbitrary function.
	
	Hence, we have  that this system admits the most general extension, with $c$ positive or negative.

	\item We consider now
	
	\begin{equation}
	L=p^4+f(q)p^2+V(q),
	\end{equation}
	If we assume $G=g(q)p$ and solve the coefficients of the monomials in $p$ in (\ref{e1}) equal to zero, we obtain two solutions, one
	\begin{eqnarray}
	V&=& \frac 1{4}(\frac 1{16}q^2c+\frac 1{8C_1}cC_2q-\frac 12\frac{C_3}{C_1(C_1q+C_2)^2}+C_4)^2-\frac {c_0}c,\cr
	g&=& C_1q+C_2, \cr
	f&=& \frac 1{16}q^2c+\frac 1{8C_1}cC_2q-\frac 12\frac{C_3}{C_1(C_1q+C_2)^2}+C_4,
	\end{eqnarray}
	that, substituted in $L$, gives, by assuming $c\neq 0$,
	\begin{eqnarray}\label{psq}
	L&=&\frac 1{1024C_1^2(C_1q+C_2)^4}\left(32p^2(C_1^3q^2+2C_1^2C_2q +C_1C_2^2)+C_1^3cq^4\right.\cr &+&\left.4C_1^2C_2cq^3+16C_1^3C_4q^2+5C_1C_2^2cq^2+(32C_1^2C_2C_4+2C_2^3c)q\right.\cr
	&+&\left.16C_1C_2^2C_4-8C_3\right)^2-\frac {c_0}c,
	\end{eqnarray}
	
	and the other, holding for $c\neq 0$ also
	
	\begin{eqnarray}
	V &=& \frac 1{(C_1q+C_2)^4}\left(C_4-\frac 1{1024cC_1^3}q(-C_1^7c^3q^7-8C_1^6C_2c^3q^6\right.\cr
	&-& 28C_1^5C_2^2c^3q^5-56C_1^4C_2^3c^3q^4+(-70C_1^3C_2^4c^3+1024c_0C_1^7\cr
	&-& 32C_1^5C_3c^2)q^3+(4096c_0C_1^6C_2-128C_1^4C_2C_3c^2-56C_1^2C_2^5c^3)q^2\cr
	&+&(-28C_1C_2^6c^3+6144c_0C_1^5C_2^2-192C_1^3C_2^2C_3c^2)q-8C_2^7c^3\cr
	&+&\left.4096c_0C_1^4C_2^3-128C_1^2C_2^3C_3c^2)\right),\cr
	g&=& C_1q+C_2,\cr
	f&=& \frac 1{16C_1^2}(C_1q+C_2)^2c+\frac {C_3}{(C_1q+C_2)^2},
	\end{eqnarray}
	where the $C_i$ are constants. It is interesting to remark that in the last case $L$ is not in general a perfect square plus a constant, as in (\ref{sq}) and (\ref{psq}).

\item We assume now 
\begin{equation}
L=\left(p_1^2+\frac 1{(q^1)^2}p_2^2+V(q^1,q^2) \right)^2,
\end{equation}
that is the square of a natural Hamiltonian on $\mathbb E^2$, and search for functions $V$ allowing the existence of non-trivial solutions $G$ of (\ref{e1}). Again, by assuming $G(q^1,q^2)$ and  collecting the  terms in $(p_1,p_2)$ in (\ref{e1}), the requirement that the coefficients of the momenta are identically zero, after assuming $c_0=0$,   gives the following solution 
\begin{eqnarray}
G&=& \left(\sin(q^2)C_2+\cos(q^2)C_3\right)q^1+C_1,\cr
V&=&-\frac c8 \frac{(C_3\sin(q^2)-C_2\cos(q^2))^2(2\tan(q^2)C_2C_3-C_2^2+C_3^2)}{C_3^2(\tan(q^2)C_3-C_2)^2} (q^1)^2\cr
&+&\frac c4\frac{(C_3\sin(q^2)-C_2\cos(q^2))C_1}{C_3(\tan(q^2)C_3-C_2)}q^1\cr
&+&F\left((\sin(q^2)C_3-\cos(q^2)C_2)q^1\right),
\end{eqnarray}
where $F$ is an arbitrary function. 


\end{enumerate}

In all the examples above all the elements of the extension procedure are polynomial in the momenta, therefore, the extended Hamiltonian and its characteristic first integral are globally defined in the same way as for the natural Hamiltonian case.

\section{Extensions of  the Two Point-Vortices Hamiltonian} 
        	 
 The dynamics of two point-vortices $z_j=x_j+iy_j$  of intensity $k_j$, $j=1, 2$, in a plane $(x,y)$ is described, in canonical coordinates $(Y_i=k_iy_i,X_i=x_i)$, by the Hamiltonian 
 $$
 L=-\alpha k_1k_2\ln \left((X_1-X_2)^2+\left(\frac{Y_1}{k_1}-\frac{Y_2}{k_2}\right)^2\right),
 $$
 where $\alpha=\frac 1{8\pi}$ and $k_i$ are real numbers \cite{TT}.
 
 If $k_2\neq -k_1$, the functions  $( k_1z_1+k_2z_2, L)$ are independent first integrals of the system (three real functions). If $k_2=-k_1$, the functions  above give only two real independent first integrals.
 
 The coordinate transformation
 $$
 \tilde X_1=(X_1-X_2)/2,\quad  \tilde X_2=(X_1+X_2)/2,\quad  \tilde Y_1=Y_1-Y_2, \quad  \tilde Y_2=Y_1+Y_2,
 $$
 is canonical and transforms  $L$ into
$$
L=-\alpha k_1k_2\ln \left(4\tilde X_1^2+\left(\frac {\tilde Y_1 +\tilde Y_2}{2k_1}  +\frac {\tilde Y_1 -\tilde Y_2}{2k_2} \right)^2\right).
$$

  The exension of $L$ is possible if global solutions $G$ of 
  (\ref{e1}) are known. We consider below two cases
  
\begin{enumerate} 
 \item If $k_1=k_2=k>0$ the Hamiltonian becomes
 $$
 L=-\alpha k^2 \ln \left(4\tilde X_1^2+\frac {\tilde Y_1^2}{k^2}\right).
 $$

 For $c=0$, $G$ in this case can be computed by using Maple, obtaining
\begin{eqnarray}
G= \left(\frac {\tilde Y_1+2ik\tilde X_1}{\sqrt{Q_1}}\right)^{\frac{Q_1\sqrt{2c_0}}{4\alpha k^3} } F_1 +  \left(\frac {\tilde Y_1+2ik\tilde X_1}{\sqrt{Q_1}}\right)^{-\frac{Q_1\sqrt{2c_0}}{4\alpha k^3} } F_2,
\end{eqnarray}
where $Q_1=k^2e^{-\frac{L}{\alpha k^2}}=4\tilde X_1^2+\frac {\tilde Y_1^2}{k^2}$ and  $F_i$ are arbitrary functions of $L$.

The function $G$ is not single-valued in general, but it is, for example, when $\frac{Q_1\sqrt{2c_0}}{4\alpha k^3}$ is an  integer. 

Let us consider now  $X_LG$, since $Q_1$ depends on the canonical coordinates through $L$ only, $Q_1$ and the exponents in $G$ behave as constants under the differential operator $X_L$, hence, the exponents in it remain integer if they are integer in $G$ and $X_LG$ is well defined on $T^*M$. Therefore, both $H$ and its characteristic first integral are globally well defined for integer values of $\frac{Q_1\sqrt{2c_0}}{4\alpha k^3} $. 

We have in this case an example in which the possibility of finding an extension depends on the parameters of the system and, in particular, on the values of the constant of motion  $L$.

\item If $k_2=-k_1=-k$, $k>0$ the Hamiltonian is
$$
L=\alpha k^2 \ln \left(4\tilde X_1^2+ \frac{\tilde Y_2^2}{k^2}\right).
$$
For $c=0$, the solution $G$ is obtained by Maple as 
\begin{eqnarray}
G= F_1\sin \left(\frac{\sqrt{2c_0}Q_2 \tilde X_2}{2\alpha k^2 \tilde Y_2}\right) +   F_2 \cos \left(\frac{\sqrt{2c_0}Q_2 \tilde X_2}{2\alpha k^2 \tilde Y_2}\right),
\end{eqnarray}   
where $Q_2=k^2e^{\frac{L}{\alpha k^2}}$ and $F_i$ are arbitrary first integrals of $L$.

It is evident that the function $G$  above, real or complex, is always globally defined, as well as $X_LG$, up to lower-dimensional sets, and this  makes possible the effective extension of the Hamiltonian $L$.

We observe that in this case the extended Hamiltonian has four independent constants of motion.

\end{enumerate}

\section{Hamiltonians with no known extension}

The procedure of extension can be applied in any Poisson manifold, not only in symplectic manifolds with canonical symplectic structure, as in the examples  above. 
Indeed, if $\pi$ is the symplectic form or the Poisson bivector determining the Hamiltonian structure of the system of Hamiltonian $L$ in coordinates $(x^1, \ldots,x^n)$, then the  symplectic or Poisson structure of the extended manifold in coordinates $(u,p_u,x^1, \ldots,x^n)$ is given by
	$$
	\Pi=\left(\begin{array}{cc|c} 
	0 & 1 &  0 \\
	-1 & 0  &  \\
	\hline
0	&   &   \pi  
	\end{array}
	\right).
	$$

 We recall that the Hamiltonian vector field of $L$ on Poisson manifolds with Poisson vector $\pi$ is $\pi dL$.

We consider below two cases of Hamiltonian systems for which we are unable to find extended Hamiltonians. The obstruction to the extension lies in both cases in the  non-global definition of the known solutions of (\ref{e1}).

\subsection{The Lotka-Volterra system}

It is well known that the Lotka-Volterra prey-predator system 
\begin{eqnarray}
\dot x=a x-b x y\\
\dot y=d xy -gy,
\end{eqnarray}
where $a,b,d,g$ are real constants, can be put in Hamiltonian form  (see  \cite{Nu}), for example with Poisson bivector
\begin{equation}
\pi=\left ( \begin{matrix} 0 & A \cr -A & 0 \end{matrix} \right),\quad A=-x^{1+g}y^{1+a}e^{-by-dx},
\end{equation}
and Hamiltonian
\begin{equation}
L=x^{-g}y^{-a}e^{dx+by}.
\end{equation}
Since the manifold is symplectic, there is only one degree of freedom, and the existence of the Hamiltonian itself makes the system superintegrable.
The  equation (\ref{e1}) with $c=0$ admits solution $G$ of the form
\begin{equation}
G=F_1(L) e^{-B} +F_2(L) e^B,
\end{equation}        	
where 
$$
B=-\frac{\sqrt{-2c_0}}a\int{\left[t\left(W \left( -\frac ba t^{-\frac ga}x^{\frac ga}ye^{\frac{d(t-x)-by}a}\right)+1\right)\right]^{-1}dt},
$$
and $W$ is the Lambert W function, defined by
$$
z=W(z)e^{W(z)}, \quad z\in \mathbb C.
$$
If we put $F_1=\frac 12 \left(\alpha+\frac \beta i\right)$, $F_2=\frac 12 \left(\alpha-\frac \beta i\right)$, where $\alpha$ and $\beta $ are real constants, then 
$$
G=\alpha \cos B - \beta \sin B.
$$
However, the Lambert W function is multi-valued in $\mathbb C-\{0\}$, even if its variable is real (in this case, it is defined only for $z\geq-1/e $ and double-valued for $-1/e<z<0$). Therefore, such a $G$ cannot provide a globally defined first integral and  does not determine an extension of $L$.  

By comparison, the Hamiltonian of the one-dimensional harmonic oscillator admits iterated extensions with $c=0$ and $c_0$ always equal to the elastic parameter of the first oscillator. One obtains in this way the $n$-dimensional, $n\in \mathbb N$, anisotropic oscillator with parameters having rational ratios, and therefore always superintegrable \cite{CDRraz}. This is not the case for the Lotka-Volterra system, where the periods of the closed trajectories in $x,y$ are not  all equal, as happens for the harmonic oscillators.

\subsection{The Euler system}

It is well known that the Euler rigid-body system is described by the Hamiltonian
$$
L=\frac 12\left(\frac {m_1^2}{I_1}+\frac {m_2^2}{I_2}+\frac {m_3^2}{I_3}\right),
$$
on the Poisson manifold of coordinates $(m_1,m_2,m_3)$ and Poisson bivector
$$
\pi=\left(\begin{matrix}0 & -m_3 & m_2 \cr m_3 & 0 &-m_1 \cr -m_2 & m_1 & 0 \end{matrix}   \right).
$$
The $(m_i)$ are the components of the angular momentum in the moving frame and they are conjugate momenta of the three components of the principal axes along one fixed direction.

A Casimir of $\pi$  is 
$$
M=m_1^2+m_2^2+m_3^2.
$$
The system has two functionally independent constants of the motion: $L$ and one of the components of the angular momentum in the fixed frame.

A solution of  equation (\ref{e1}) can be found by using the Kuru-Negro \cite{CRsl} ansatz
\begin{equation}\label{KN}
X_L G=\pm \sqrt{-2(cL+c_0)}G,
\end{equation}
whose solutions are solutions of the  equation (\ref{e1}) too.
A solution of (\ref{KN}) is
\begin{eqnarray}
G=f e^{\left[\mp \frac{I_1I_2I_3}{\sqrt{I_2(I_1-I_3)}}\sqrt{\frac{-2(cL+c_0)}{X_2}}F\left(m_1\sqrt{\frac{I_2(I_1-I_3)}{X_1}},\sqrt{\frac{I_3(I_1-I_2)X_1}{I_2(I_1-I_3)X_2} }\right)\right]},
\end{eqnarray}
where $f$ is an arbitrary function of the first integrals $L$ and $M$,
\begin{eqnarray}
X_1=I_1I_2(M-2I_3L),\\
X_2=I_1I_3(2I_2L-M),
\end{eqnarray}
and $F(\phi,k)$ is the incomplete elliptic integral of first kind 
$$
F(\phi,k)=\int_0^\phi \frac {d\theta}{\sqrt{1-k^2\sin ^2 \theta}},
$$
which is a multiple valued function, being the inverse of Jacobi's sinus amplitudinis $sn$ function.
The function $G$, possibly complex valued and with singular sets of lower dimension, depends essentially on $m_1$, since all other arguments in it are either constants or first integrals. 

Since our function $G$ is not single-valued, we cannot in this case build an extended Hamiltonian from $L$.


\section{Conclusions}

By the examples discussed in this article we see that extended Hamiltonians can be obtained also from non-natural Hamiltonians $L$ and not only from the natural ones. The case when $L$ is quartic in the momenta is very much similar to the quadratic cases studied elsewhere, and the extension procedure does not encounter new problems.
For the two point-vortices Hamiltonian, we have that global solutions of (\ref{e1}) can be obtained in correspondence of particular choices of parameters in $L$.
In the remaining examples, we are unable to obtain globally defined solutions of (\ref{e1}) and we cannot build extensions in these cases.
In future works, the extended Hamiltonians obtained here could be studied in more details, while the search for global solutions for the cases of Lotka-Volterra and of the rigid body, or for the reasons of their non-existence, could be undertaken.

\

\

{\bf Conflict of Interest}: the authors
declare that they have no conflicts of interest.

\end{document}